\documentstyle[aps,prl,epsfig]{revtex}
\twocolumn
\newcommand{\Hm}{H^{\rm melt}_{\rm c}}

\newcommand{\bis}{ Bi$_2$Sr$_2$CaCu$_2$O$_{8+\delta}$}
\begin{document}
\twocolumn[\hsize\textwidth\columnwidth\hsize\csname@twocolumnfalse\endcsname
\draft
\title{Anisotropy of Vortex-Liquid and Vortex-Solid 
Phases in Single Crystals of Bi$_2$Sr$_2$CaCu$_2$O$_{8+\delta}$: 
Violation of the Scaling Law}

\author{J. Mirkovi\'c\cite{byline}, S. E. Savel'ev, E. Sugahara, and K. Kadowaki}
\address{Institute of Materials Science,
         University of Tsukuba, Tsukuba 305-8573, Japan}
\maketitle
\begin{abstract}
The vortex-liquid and vortex-solid phases in single crystals of
Bi$_2$Sr$_2$CaCu$_2$O$_{8+\delta}$ placed in tilted magnetic
fields are studied by 
in-plane resistivity measurements using the Corbino geometry to avoid
spurious surface barrier effects. 
It was found that the anisotropy of the vortex-solid phase increases with temperature and
exhibits a maximum at
$T\approx 0.97\ T_c$.
In contrast, the anisotropy of the vortex-liquid 
rises monotonically across the whole measured temperature range. The
observed behavior is discussed in the context of dimensional crossover 
and thermal fluctuations of vortices in the strongly 
layered system. 
\end{abstract}
\tighten

\newpage
\vskip.2pc]


The anisotropic static and dynamical properties of superconductors  
have commonly been described by the three-dimensional Ginzburg-Landau (hereafter 3DGL)
theory with a single parameter 
of anisotropy $\gamma$ defined as a ratio $(m_c/m_{ab})^{1/2}$ 
 of the effective masses $m_c$ and $m_{ab}$ along the $c$-axis and in the 
$ab$-plane, respectively \cite{bible}.  The anisotropy is related to the basic 
parameters of a superconductor as 
$
\gamma=\xi_{ab}/\xi_c=\lambda_c/\lambda_{ab}=
H_{c2\parallel}/H_{c2\perp},
$
where $\xi_{ab}$ and $\xi_{c}$ are the in-plane and the out-of-plane coherence 
lengths, $\lambda_c$ and $\lambda_{ab}$ are the magnetic field penetration 
depths  along the $c$-axis and in the $ab$-plane, while 
$H_{c2\perp}$ and $H_{c2\parallel}$ are the out-of-plane and the in-plane upper critical 
magnetic fields.

Based on 3DGL, Blatter {\it et al.} \cite{scale} obtained the general 
scaling law, which 
can be applied for the description of different physical properties 
of anisotropic superconductors 
in oblique magnetic fields. For instance, 
the resistivity in a magnetic field $H$, tilted away 
from the $c$-axis for the angle $\theta$, depends only on the 
reduced field as
\begin{equation}
\rho(H,\theta)=\rho(H\sqrt{\cos^2\theta+\sin^2\theta/\gamma^2}).
\label{rho}
\end{equation}
The scaling law for the vortex lattice melting 
magnetic field was derived in a similar manner as  
\begin{equation}
H^{melt}(\theta)=H^{melt}(\theta=0)/\sqrt{\cos^2\theta+\sin^2\theta/\gamma^2}.
\label{melt}
\end{equation} 

In previous studies it was found
 that the scaling law describes  well experimental data
 such as resistivity \cite{YBCO,kwok}, magnetization \cite{welp}, and thermodynamic
 measurements \cite{schilling} in ${\rm YBa_2Cu_3O_{7-\delta}}$ with anisotropy
parameter $\gamma\approx 5-7$ \cite{bible}. 
 Concerning the higher anisotropic layered systems,
 it was recognized earlier that the dissipation 
in Bi$_2$Sr$_2$CaCu$_2$O$_{8+\delta}$ depends practically only on 
the $c$-axis magnetic field component \cite{kes,iye} 
except very close to the $ab$-plane 
\cite{roas} which is in agreement with the 3DGL scaling law with high anisotropy
parameter. However, it was reported \cite{naughton} that the resistivity data near 
the $ab$-plane agree better with Thinkam's thin film (hereafter 2DT) model \cite{tinkham}, 
than with the 3DGL theory. Later, it was also found that the measured 
anisotropy parameter for Bi$_2$Sr$_2$CaCu$_2$O$_{8+\delta}$ thin films increases with temperature, 
according to the 2DT model, followed by an 
indication of the two to three-dimensional (2D-3D) 
crossover near the transition temperature $T_c$ \cite{silva1,silva2}.  On the other 
hand,  recent experimental studies of the first-order vortex-lattice melting 
transition (hereafter VLMT) have shown a linear decay of the out-of-plane melting field
with an increasing in-plane magnetic field, even near the $c$-axis \cite{oi,mi}, in a
strong contrast to the above mentioned approaches. 
In a strict sense, it is reasonable to expect that 
the scaling law based on the  3DGL model may not be applicable to
the extremely anisotropic layered system 
such as Bi$_2$Sr$_2$CaCu$_2$O$_{8+\delta}$. In this material, the coherence length along 
the $c$-axis at low temperatures is smaller than the lattice 
parameters of the crystal, which means that
the layerness of the material has to be 
considered \cite{bible}. 
Thus, the 3DGL scaling approach \cite{scale},
 based on a continuous medium,   
 is not valid due to the discreteness of the system, and therefore,
it is not at all trivial that 
the expressions (\ref{rho}) and (\ref{melt}) are still valid. 

Hence, there is still a question of how the anisotropy  
 manifests itself in the different vortex phases in \bis \ single crystals.  
In this Letter, we report the comparative study of the anisotropy
of the vortex-solid and vortex-liquid phases in single crystals of 
Bi$_2$Sr$_2$CaCu$_2$O$_{8+\delta}$.
The anisotropy of the first-order vortex lattice melting transition  
increases with temperature in the range of $T<0.97\ T_c$, 
whereas at higher temperatures, the VLMT anisotropy decreases. 
In contrast to the anisotropy of the VLMT, the anisotropy of the in-plane
resistivity in the vortex liquid phase increases monotonically with temperature 
in the whole measured range. 

 The in-plane resistivity measurements were performed
for three as-grown single crystals of $\rm Bi_2Sr_2CaCu_2O_{8+\delta}$
with transition temperatures $T_c^{(I)}= 90.3$ K, $T_c^{(II)} = 86.0$~K, and $T_c^{(III)}= 90.0$ K, for the samples \#s1, \#s2 and \#s3,
respectively.   In order to probe the true bulk 
properties, {\it i.e.,} 
to avoid the surface barrier effects which occur in  
conventional strip geometry \cite{fuchs}, we have 
used the Corbino
electrical contact configuration (see inset in Fig.~1a).  
The diameters of
the Corbino discs were $D^{(I)}=1.9$ mm, $D^{(II)}=1.95$ mm, and
$D^{(III)}=2.7$ mm while the thickness was $t\approx 20\ \mu$m for
all samples.  The resistance has been measured by using the ac lock-in 
technique at a low frequency of 37 Hz. 
The magnetic 
field, generated up to 70 kOe by a superconducting split 
\begin{figure}[btp]
\begin{center}\leavevmode
\includegraphics[width=1.15\linewidth]{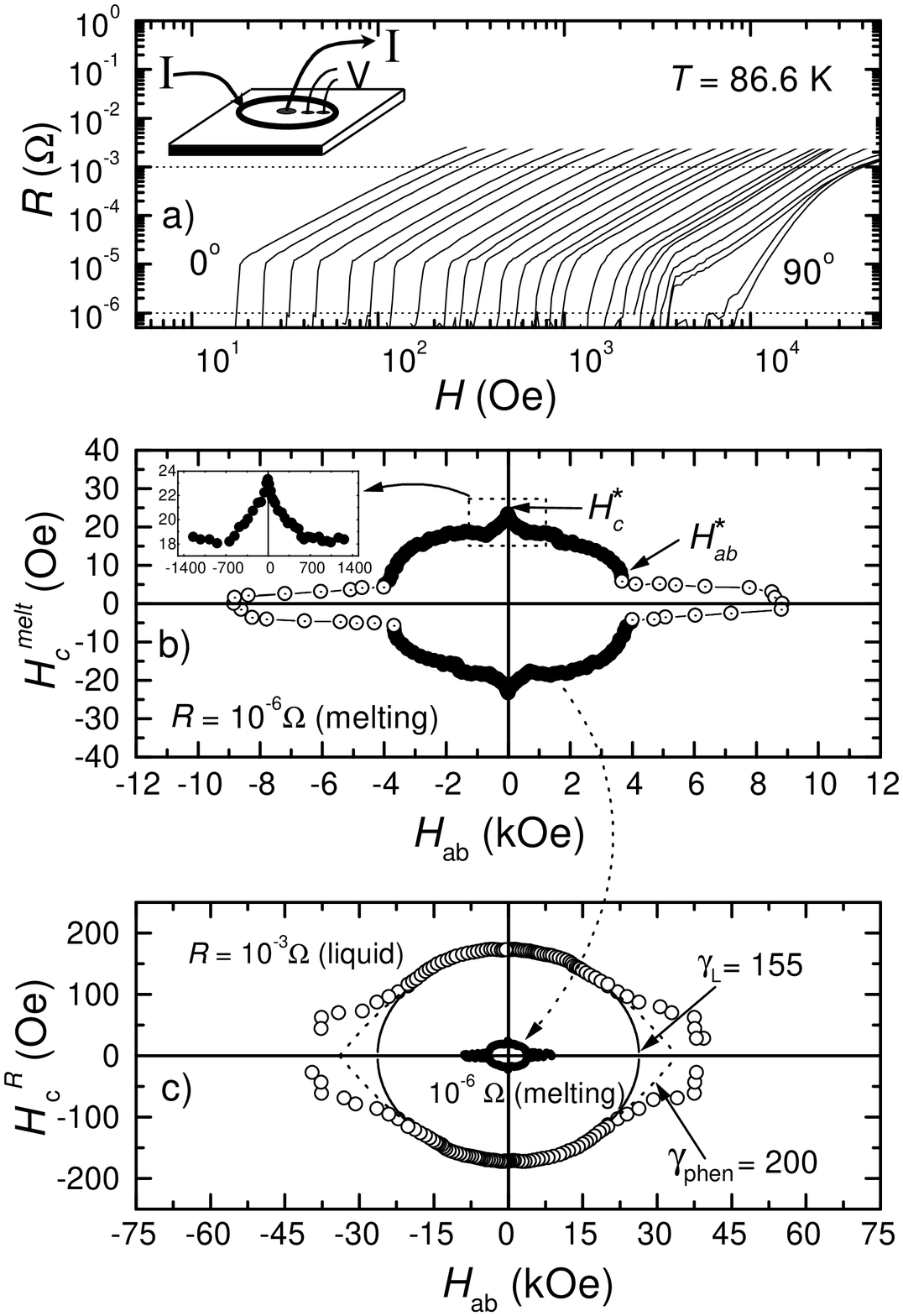}
\caption{ (a) The magnetic field dependence of the resistance at various field orientations 
from ${\rm \bf H}\parallel {\rm\bf c}$ ($0^\circ$) to ${\rm \bf H}\parallel 
{\rm\bf ab}$($90^\circ$)
at $T$=86.6 K (sample \#s1). Inset in (a): electrical contacts in the 
Corbino geometry.
(b) The $H_c-H_{ab}$ phase diagram of the vortex lattice 
melting transition at $T=86.6$ K: 
the filled symbols mark 
the first-order phase transition, while the open ones correspond to the 
continuous resistivity transition. Inset in (b): the part of the VLMT phase diagram 
near the $c$-axis, plotted at a magnified scale.
(c) The equi-resistance contour at 86.6 K
(open symbols) in the vortex-liquid phase. The
 melting transition from (b) is replotted at the centre. 
The solid line is the ellipse with anisotropy value of $\gamma_L=155$, 
while the dotted line corresponds to the phenomenological 
formula wih $\gamma_{phen}=200$.} 
\label{f1}
\end{center}
\end{figure}
\noindent
 magnet, was rotated by a fine goniometer with angular resolution of
$0.01^\circ$.

A typical set of the resistance curves is presented in Fig.~1a for $T=86.6$ K.
The sharp resistance drop, attributed 
to the VLMT \cite{watauchi}, is clearly detected 
 across the  wide angular range ($\theta<89.86^\circ$) \cite{mi}. 
Using the resistance criteria of $R=10^{-6}\Omega$,
 which is set somewhat below the resistivity kink, and 
 $R=10^{-3}\Omega$, which is well above the VLMT anomaly,
 we construct the phase diagram of the VLMT and
the equi-resistance contour for the vortex-liquid in the $H_c-H_{ab}$ phase plane (Fig. 1(b,c)). 
The VLMT phase line $\Hm(H_{ab})$ (Fig. 1b)
exhibits the peculiar step-wise shape \cite{mi}.
 Namely, starting from the $c$-axis, the out-of-plane component of the VLMT, $\Hm$, decays 
linearly with increasing $H_{ab}$ \cite{oi}, which is associated 
with the crossing vortex lattice \cite{kosh}. 
As the magnetic field further approaches the $ab$-plane, the linear 
dependence $\Hm(H_{ab})$ sharply transforms to the plateau (inset in Fig.1 b), which is
attributed to the trapping of pancake vortices by Josephson vortices \cite{prb-mi}.
Very close to the $ab$-plane ($\theta\approx 89.96^\circ$), the
first-order phase transition (filled symbols in Fig. 1b) 
changes to the 
continuous resistance transition (open symbols), which exhibits 
the cusp in the $H_c-H_{ab}$ phase plane \cite{mi}.
In strong contrast to the linear decay of $\Hm(H_{ab})$ around the $c$-axis, 
the equi-resistance line $H^R_c(H_{ab})$ 
(Fig. 1c) in the vortex-liquid phase demonstrates smooth quadratic behavior. 
 Near the $ab$-plane, the equi-resistance contour still has 
a tail, as if it traces the VLMT cusp. The cusp-like shape of both the VLMT line and 
the equi-resistance contour, pronounced near the $ab$-plane,  
is possibly associated with the intrinsic pinning \cite{tachiki} 
which, surprisingly, seems to be active even in the vortex 
liquid phase, quite far from the VLMT.   

The next question is how to estimate the anisotropy of the 
vortex liquid phase and the VLMT. 
The smooth quadratic dependence $H^R_c(H_{ab})$, can be derived from the general 
symmetry law $\rho(H_c,H_{ab})=\rho(H_c,-H_{ab})$, assuming that the resistivity
is an analytical function of $H_{ab}$ in the vortex-liquid.  
The simple analysis \cite{aniz} gives the angular dependence  
$H_c^R(H_{ab})=H_c^R(0)-H_{ab}^2/(H_{c}^R(0)\gamma_{phen}^2)$ 
for the limited angular range $0^\circ<\theta<90^\circ-57.3^\circ/\gamma_{phen}$
 with phenomenological anisotropy parameter $\gamma_{phen}(T,H_c)$. 
This equation succesfully fits our data at 86.6~K in the vortex liquid phase
except very close to
the $ab$-plane with $\gamma_{phen}=200$ (dotted line in Fig. 1c). 
It is worth noting that the phenomenological 
angular dependence $H_c^R(H_{ab})$ coincides with the angular dependence
of the upper critical field $H_{c2}(\theta)$ for thin superconducting films 
obtained by Tinkham \cite{tinkham} as
$H_{c2}(\theta)\cos\theta/H_{c2\perp}+H_{c2}^2(\theta)\sin^2\theta/{\gamma^2_{2DT} 
H^2_{c2\perp}}=1$
with the field independent anisotropy parameter $\gamma_{2DT} \propto 1/\sqrt{1-T/T_c}$.
On the other hand, to describe the equi-resistance contours in the vortex liquid phase,
it is convenient to apply the more common 3DGL model. 
Following the 3DGL theory, the $H_c^R(H_{ab})$ line should be an 
ellipse $(H_c^R(H_{ab}))^2+H_{ab}^2/\gamma_{3DGL}^2=(H_c^R(0))^2$ with anisotropy 
$\gamma_{3DGL}$ being 
independent of magnetic field and temperature. 
Fitting the curve in Fig. 1c to 
an 
ellipse with an anisotropy parameter of 155, we also find a reasonable 
agreement only in a 
limited angular range $\theta<89.75^\circ$. 
Moreover, \hfill the empirical \hfill anisotropy $\gamma_L$ of the vortex 
liquid, \hfill obtained \hfill by \hfill the  \hfill
elliptical \hfill fitting \hfill of \hfill the \hfill
equi-
\begin{figure}[btp]
\begin{center}\leavevmode
\includegraphics[angle=-90,width=1.15\linewidth]{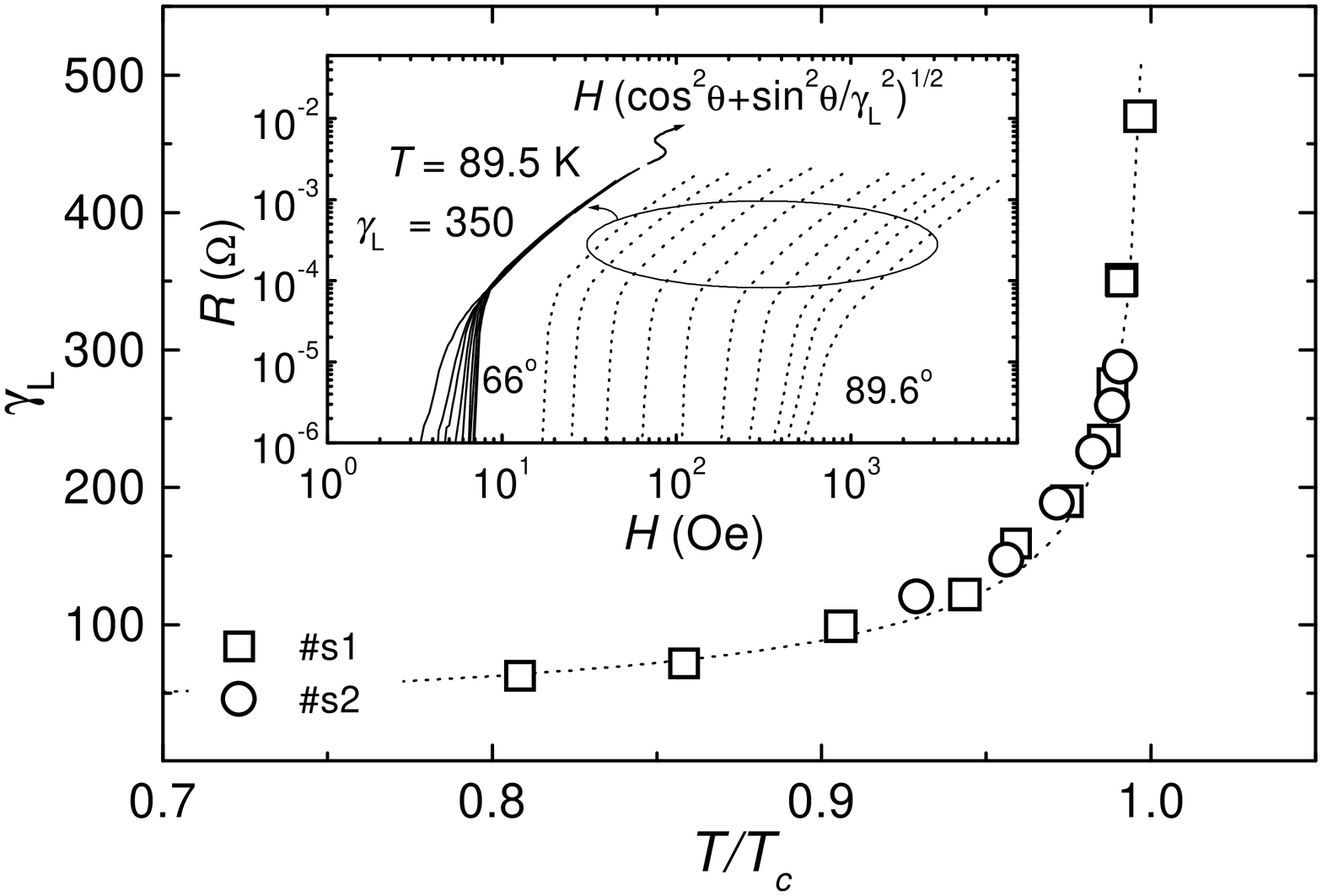}
\caption{ The temperature dependence of $\gamma_L$, extracted by  
fitting the equi-resistance lines to an ellipse, for samples
\#s1 and \#s2. The dashed line corresponds to the dependence 
$\gamma_L=28/(1-T/T_c)^{1/2}$.
Inset: The dotted curves show the $R(H)$ data obtained at $T$=89.5 K
at various magnetic field 
orientations with respect to the $c$-axis (from left, $\theta=66^\circ,
74^\circ, 80^\circ, 84^\circ, 86.5^\circ, 88.05^\circ, 88.75^\circ, 89.2^\circ,
89.39^\circ, 89.6^\circ$). The solid curves represent scaled
resistance curves according to equation (\ref{rho}) with $\gamma=\gamma_L=350$.} 
\label{f2}
\end{center}
\end{figure}
\noindent
resistance
contours, 
increases strongly with temperature in 
the whole measured temperature interval (Fig. 2), which
indicates that the  
3DGL scaling law is violated.  Thus, the anisotropy 
parameter $\gamma_L$ extracted from the ellipse can no longer be attributed to the 
ratio of the effective masses as assumed in the 
3DGL theory and may reflect the two-dimensional character of the vortex system in analogy to 
the 2DT approach \cite{tinkham}.   Interestingly, 
the resistivity curves above the VLMT kink, 
measured at different field orientations 
($\theta<89.6^\circ$, $T=89.5$ K, $R<2\cdot 10^{-3}\Omega$), 
collapse into a unique one if the curves 
$R(H,\theta)$ are scaled in a common way by using eq.~(\ref{rho}) 
(see inset in Fig. 2) with $\gamma=\gamma_L$. 
Hence, 
the anisotropy depends weakly on the resistance level in the vortex liquid
but changes sharply around the vortex-lattice melting transition (see Inset in 
Fig. 2), implying different dimensionality of the probed vortex phases.  

Recognizing the fact that the VLMT does not follow the 3D scaling 
law (\ref{melt}), 
the effective anisotropy of the 
VLMT can be estimated by the ratio $\gamma_{melt}=H_{ab}^*/H_c^*$ 
(for definition of
$H_c^*$ and $H_{ab}^*$, see Fig. 1b).  
It is necessary to emphasize that the VLMT phase lines, obtained at 
different temperatures, 
fall roughly on a unique line if plotted 
in the phase plane $H_c/H_c^*-H_{ab}/H_{ab}^*$, 
as shown in Fig. 3a. Therefore,
the whole VLMT line in the 
$H_c-H_{ab}$ plane changes proportionally with temperature, 
 which justifies the chosen definition of the anistropy of the \hfill VLMT.  \hfill
The \hfill temperature \hfill dependences \hfill of \hfill the 
\begin{figure}[btp]
\begin{center}\leavevmode
\includegraphics[width=1.15\linewidth]{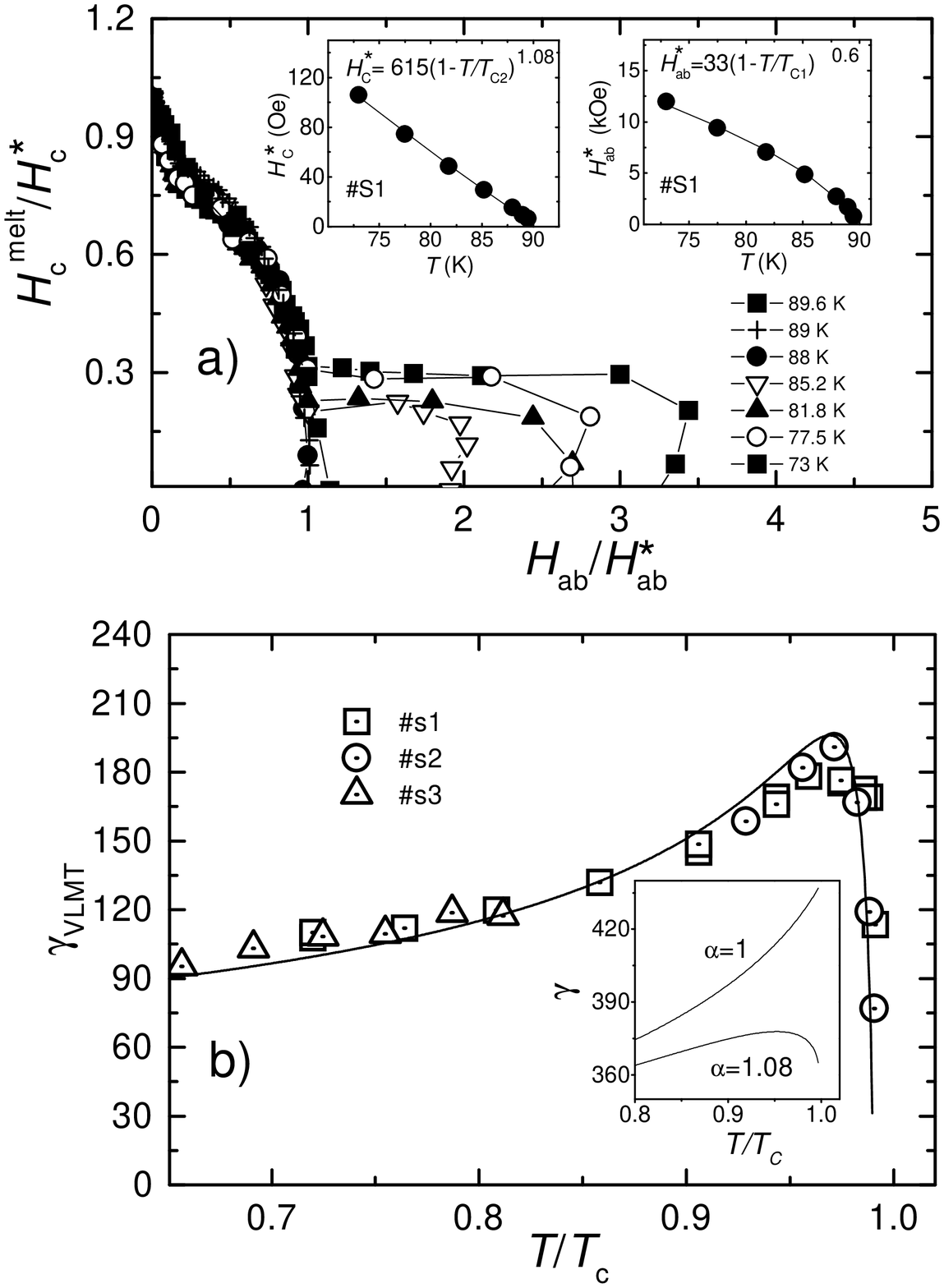}
\caption{(a) The phase diagrams of the vortex melting transition plotted in the phase plane 
$H_c/H_c^*-H_{ab}/H_{ab}^*$ at different temperatures for \#s1. 
The left and right insets display the temperature dependence of $H_{ab}^*$ and $H_{c}^*$, 
respectively (for definition of the fields $H_{ab}^*$ and $H_{c}^*$, see Fig. 1b).
(b) The temperature dependence of the VLMT anisotropy 
defined as $\gamma_{melt}=H^*_{ab}/H^*_{c}$, obtained for three samples. 
The solid line is the empirical equation for $\gamma_{melt}(T)$ discussed in the text.
Inset: the calculated temperature dependence of the anisotropy
in the frame of the model [23].} 
\label{f3}
\end{center}
\end{figure}
\noindent
characteristic
fields $H_c^*$ and $H_{ab}^*$
exhibit different curvatures (see insets in Fig. 3a), which 
result in a maximum of 
the temperature dependence of the VLMT anisotropy  $\gamma_{melt}(T)$ at 
$T\approx 0.97\ T_c$ as it is seen in Fig 3b. 
The obtained temperature dependence of $\gamma_{melt}$
may be approximated by the empirical equation 
$\gamma_{melt}(T)=\gamma_0(1-T/T_{c1})^{\beta}/(1-T/T_{c2})^{\alpha}$
(the solid line in Fig. 3b)
with parameters $T_{c1}=89.4$ K, $T_{c2}=90.7$ K, $\alpha=1.08$ and $\beta=0.6$
obtained by fitting the fields $H_c^*(T)=H_c^*(0)(1-T/T_{c2})^{\alpha}$ and
$H_{ab}^*(T)=H_{ab}^*(0)(1-T/T_{c1})^\beta$ for the sample \#s1. 
The two characteristic temperatures ($T_{c1}$, $T_{c2}$) 
in the above empirical formula, may appear due to the fact that
the thermodynamical vortex fluctuations have different influence on 
the interlayer and the intralayer-superconductivity in the 
 tilted magnetic fields. 
The similar maximum of $\gamma(T)$, accompanied by
the disappearance of the cusp in the dependence $H_{onset}(\theta)$ 
($R(H_{onset})=0$) 
near the $ab$-plane, was observed 
earlier in thin films of Bi$_2$Sr$_2$CaCu$_2$O$_{8+\delta}$ \cite{silva1},
although there was neither indication of the VLMT nor clear evidence 
which vortex phase was probed. 
The phenomenon \cite{silva1} was interpreted \cite{silva2} 
as a crossover from 2D to 3D behavior, which occurs
 when the coherence length $\xi_c$ exceeds the distance between 
CuO$_2$ layers. In agreement with this scenario, 
the cusp in the $H_c-H_{ab}$ phase diagram, 
associated with  the continuous resistivity transition,  
also disappears around the temperature (see Fig. 3a)
at which $\gamma_{melt}(T)$ is at a maximum \cite{physc}.  
However, based on the 2D-3D crossover, 
it is impossible to explain why the anisotropy of the vortex liquid phase still 
increases at temperatures $T>0.97\ T_c$ where $\gamma_{melt}(T)$ decreases. 
Moreover, the 2D-3D scenario proposed in \cite{silva2} describes only
the temperature and angular dependence of the upper critical field 
$H_{c2}(\theta,T)$, which, strictly speaking, can not be applied to
the VLMT and the equi-resistance contours. 

Another possible origin of the temperature dependence of the anisotropy could be related 
to the thermal fluctuations of vortices, which suppress the Josephson 
coupling in-between the CuO$_2$ layers more efficiently than the superconductivity within the layers.
As a result, $\lambda_c(T)$ increases faster than $\lambda_{ab}(T)$ and, thus, 
the anisotropy $\gamma=\lambda_c/\lambda_{ab}$ rises with increasing temperature. 
According to the model proposed for the magnetic field parallel to the $c$-axis
 \cite{bul}, the temperature dependence of the anisotropy is determined 
by the equation $\gamma^2/\gamma_0^2=\exp{\left(\delta(H_c,T)\gamma^2/\gamma_0^2\right)}$
\cite{ss}, where
$
\delta=\pi TH_cs\lambda^2_{ab}(T)\gamma_{0}^2/2\Phi_0^3
$
with $\gamma_0=\gamma(T=0)$. In order to take into account the temperature
dependences of the VLMT and equi-resistance magnetic fields,
 the empirical equation $H_c=H_0(1-T/T_c)^{\alpha}$ is used. 
Since $\gamma$ is a monotonical function of 
$\delta\propto T\left(1-T/T_c\right)^{\alpha-1}$,
 the temperature dependence of $\gamma$ has a maximum 
 approaching $T_c$, if $\alpha>1$, otherwise $\gamma$ monotonically  
increases with $T$. The inset in Fig. 3b
displays the temperature dependence of $\gamma$ obtained by using the above equations 
with $\gamma_0=300$ and $\alpha=1.08$, corresponding to the measured VLMT 
(inset in Fig. 3b), and $\alpha=1$, as assumed for the vortex-liquid phase. 
Nevertheless, the calculated value of 
the magnetic field $H_0=29$ kOe disagrees with the experimental value 
of the $c$-axis field component, 
but is close to the total applied magnetic field. The discrepancy
could be related to the presence of the Josephson vortices, which have not been 
considered in the model \cite{bul}. The pancake vortex sublattice interacts with the
Josephson vortex sublattice in the crossing lattice structure 
(set in the tilted magnetic fields) providing the additional 
mechanism of the renormalization of 
the anisotropy in the vortex-solid phase \cite{kosh,prb-mi}. 

In summary, in contrast to the prevailing belief, we have found 
clear experimental evidence 
 that neither the 3DGL scaling law nor the 2D Tinkham's 
model consistently describe the resistivity in the vortex-liquid phase as well as
the behavior of the vortex lattice melting transition, 
which can qualitatively be accounted for by the thermal fluctuations of vortices. 

We thank M. Tachiki, L.~N. Bulaevskii, A.~E. Koshelev, and V.~M. Vinokur
for stimulating discussions.


\begin{thebibliography}{99}
\bibitem{aniz} In a wide angular range $90^\circ-57.3^\circ/\gamma>\theta$, the equation 
$R(H_c,H_{ab})=R(H_c,0)+\nu(H_c,T) H_{ab}^2$ with 
$\nu(H_c,T)=(\partial^2 R(H_c,H_{ab})/\partial H_{ab}^2)|_{H_{ab}=0}$ 
can be used, since  a linear term does not follow the symmetry restrictions. 
Thus, $H^R_c(H_{ab})$ should satisfy the equation $R(H_c^R(H_{ab}),0)+\nu H_{ab}^2
=R(H^R_c(0), 0)$ with $R(H^R_c(0))=R^*$ and $R^*=10^{-3} \Omega$. The 
approximate solution of 
the equation, presented in the text, yields the phenomenological anisotropy 
$\gamma_{phen}(T,H_c^R(0))=\sqrt{(\partial R(H_c,0)/\partial H_c)|_{H_c=H_c^R(0)}/\nu H^R_c(0)}$.
\bibitem{ss} We use the equation obtained in Ref.~\cite{bul} for the  
Josephson coupling prevailing over the electromagnetic one 
which is valid if $\gamma s\lesssim\lambda_{ab}(T)$. The last condition is satisfied
at $T>T^*$ with $T^*/T_c=1-\lambda_{ab}^2(T=0)/\gamma^2s^2=0.8$ for $\lambda_{ab}(T=0)=2000\ \AA$,
$s=15\ \AA$, $\gamma=300$.
\end{thebibliography}
\end{document}